\documentstyle[prl,aps,epsf]{revtex}

\begin{document}
\title{Three-Dimensional Morphology of Vortex Interfaces
Driven by Rayleigh-Taylor or Richtmyer-Meshkov Instability}

\author{
N.A. Inogamov,$^{1,2,3}$
M. Tricottet,$^3$
A.M. Oparin,$^4$
and
S. Bouquet$^{1,3}$ }

\address{$^1$CEA/Saclay, Orme-des-Merisiers,
DSM/DRECAM/SPEC, 91191
Gif-sur-Yvette Cedex, France}

\address{$^2$Landau Institute for Theoretical Physics, RAS, 142432,
Chernogolovka, Moscow region,
Russian Federation}

\address{$^3$CEA/DIF, D\'epartement de Physique Th\'eorique et Appliqu\'ee,
91680 Bruy\`eres-le-Ch\^atel, France}

\address{$^4$Institute of Computer-Aided Design, RAS, 123056, Moscow,
Russian Federation}

\date{\today}

\maketitle


\begin{abstract}
We study the 3D topology of Rayleigh-Taylor (RT) and Richtmyer-Meshkov
(RM) single-modes, which includes bubbles, jets and saddle points. We
present an analytic description of the interface
as a whole, for arbitrary time-dependant acceleration $g(t).$ The
dependance of morphology on the lattice - Hexagonal, square or
triangular -of bubbles
are investigated. RM accelerations in the case of a large density
ratio produce jets well separated from each other while, in RT case,
jets are connected by liquid sheets. We compare our analytic results
to numerical simulations. 
\end{abstract}

\pacs{PACS numbers: 47.20.Ma, 47.40.Nm}

\twocolumn

The RT and RM instabilities
(RTI and RMI)
play an important role in astrophysics\cite{BruceApJSS,Serge-SN,INA},
in ion\cite{Ion} and laser\cite{HolsteinBodner}
inertial confinement fusion (ICF),
in shock-tube mixing\cite{JFMarsel,ST}
and in chemical, nuclear or thermonuclear combustion\cite{Comb}.
The asymmetry caused by RTI and RMI in spherical implosions
 strongly alters the neutron yield and energy gain
in ICF targets\cite{Ion,HolsteinBodner}.
Recently,
it was proposed
to use experiments
on very powerful existing and future laser systems
such as Omega and the National Ignition Facility in the USA
or the Laser M\'egaJoule in France
for modeling the unstable explosion
of supernovae (SN)\cite{BruceApJSS,Serge-SN}
or unstable expansion of SN-remnants\cite{BruceApJSS,Serge-SN}.
Flow with two shocks and mixing between them
(similar to SN-flow)
takes place during an expansion of detonation products
after an explosion\cite{INA,AIAA}
(compare with\cite{TychoSN}).
RTI/RMI are also significant  for other astrophysical applications
such as planetary nebulaes, Wolf-Rayet stars
and magnetospheres of neutron stars\cite{INA}.
The physical origins of exchange instabilities
(RTI and RMI)
are connected with baroclinic generation of vorticity\cite{INA,Zabusky}.
RTI is driven by buoyancy
(see reviews\cite{INA,SharpKull}).
RMI occurs after the passage of a shock wave
through surface corrugations\cite{INA,JFMarsel,ST,Zabusky}.

Configurations of 3D single-mode perturbations
can be represented by bubble lattices
having various geometrical symmetries:
hexagonal (B6), square (B4) or triangular (B3).
The B stands for ``bubble''
and the digits 6, 4 and 3
correspond to the number of bubbles
adjacent to the chosen one.
Saddles and jets also form lattices.
In the lattice B6 a jet J
has three neighbouring jets.
Therefore the lattice B6 is  the same time as the lattice J3.

Our goal is to describe 3D phenomena and their dependence on both
lattice symmetry and acceleration profile, $g(t).$ This approach is
needed to understand phenomena occuring in ICF, astrophysics, etc.,
because real unstable flows are three-dimensional
and the acceleration, $g(t),$ satisfies
neither the RTI nor the RMI conditions\cite{Logory}.
To begin our study, let us compare the examples in Fig.\ref{fig-ripples} (2D)
and Fig.\ref{fig-bubbles} (3D).
They represent,
at time\footnote{Space is normalized by wavenumber $k$ and $t$
by $k$ and initial amplitude of velocity $w_0$ (RM) or $k$ and $g$
(RT).} $t=1$ , the position
of the periodically perturbed interface $\eta$
in the RM case.
At $t=0,$
the interface $\eta(x,0)\equiv \eta_{00}$ (resp. $\eta(x,y,0)\equiv \eta_{00}$)
corresponds to the plane $z=0$
and the potential of the velocity perturbation
is $\varphi (x,z,t=0)\propto \cos x e^{-z}$
[resp. $\varphi (x,y,z,0)\propto (\cos x+\cos y) e^{-z}$]
in 2D (resp. 3D/B4).
These initial conditions
are called ``standard''
because they are widely used
for small initial perturbations.
The heavy fluid,
with density $\rho,$
is above $(z>\eta)$
the light fluid with density $\rho_l.$
Jets of dense fluid move down
while light bubbles rise up.
Analytically we consider the case with density ratio
$\mu=\rho_l/\rho\to 0.$
Our method of calculation is presented below.

In the 2D case (Fig. \ref{fig-ripples})
[resp. 3D case (Fig. \ref{fig-bubbles})],
the points
B and J
(resp. B, S and J) are very important
since they correspond to stagnation points. At these points,
the velocity of the fluid becomes zero
relative to the surface $\eta.$
      \begin{figure}[htbp]
      \caption{Rippled structure of 2D solutions. Periodic sequence of
parallel valleys. There is a  chain ..-B-J-B-J-.. of tips of bubbles B and jets J.}
      \label{fig-ripples}
      \end{figure}
      \begin{figure}[htbp]
      \caption{Geometry of a single 3D mode in a square lattice
(B4). Periodic cell structure in the horizontal plane or
two-dimensional crystal formed from points of tips of bubbles B, jets
J and saddles S marked by circles.}
      \label{fig-bubbles}
      \end{figure}

The 3D/B4 flow is invariant
relative to $90^0$-rotations around B and J vertical axis
and to $180^0$-rotations around S axis
(Fig. \ref{fig-bubbles}).

For small density ratios, $\mu\ll 1,$
the whole flow
is clearly divided
into two qualitatively very different parts.
The first is the bubble envelope
imprinting into the dense fluid.
Bubbles brake the initially continuous dense fluid
and produce jets.
Above the envelope,
the dense fluid
is still in a contiguous state. The second part is the ejecta
(jets) pinched and driven down
by the imprinting bubbles.
The points B belong to the contiguous fluid
and the points S and J to the ejecta.
The 3D pattern of the ejecta is rather complicated (Fig. \ref{fig-simul}).
The ejecta consists of
1) wall type jets
going down from B to S
and
2) leg type jets going from S to J.
The bubble has the shape of a well in the dense fluid.
This well transforms into  walls 
or skirts around the points B.

The relevant topological measure
of the form of the ejecta
is the geometrical ratio
\begin{equation}
\Gamma (t)=\Delta z_{BS}/\Delta z_{BJ},\text{ with }
\Delta z_{MN}=z_M(t)-z_N(t),
\label{gamma}
\end{equation}
where $z_B, z_S$ and $z_J$
are the vertical positions of the points (Fig. \ref{fig-simul}).
$\Gamma$ is the ratio  of the length of ``skirt''  to the length of ``legs''.

      \begin{figure}[htbp]
      \caption{Numerical results for RM (the triad in the upper panel,
      $t=20)$ and RT (lower panel, $t=9$) cases for the lattices B6,
      B4 and B3 from left to right. The ratio is $\mu=0.1.$}
      \label{fig-simul}
      \end{figure}

The six direct numerical simulations (DNS) in Fig. \ref{fig-simul}
were done by a grid-characteristics method\cite{BILAMO}.
Several interesting works
are devoted to DNS of 3D RTI/RMI flows\cite{DNS3D}.

Before presenting our results,
let us give the boundary conditions and spectral decomposition.
The motion is described by a velocity harmonic potential $\varphi$
(the vorticity is concentrated at the interface).
Classical\cite{INA,SharpKull}
kinematic and dynamic conditions
are
\begin{equation}
\eta_t=w|-\eta_x u|-\eta_y v|,\;\;
\vec v=\{u,v,w\}, \; \eta_t\equiv\partial\eta/\partial t,
\label{continuity}
\end{equation}
\begin{equation}
\varphi_t|=  \vec v^2|/2 + g(t) \eta,
\;\;\; f|\equiv f|_{\eta}\equiv f|_{z=\eta(x,y,t)}.
\label{bernouilli}
\end{equation}
We represent the potential by a Fourier series
and the surface $\eta$ near stagnation points
by a Taylor series
\begin{equation}
\varphi(x,z,t) = \sum_{n=1}^{N} \varphi_n(t) c_{nx} e^{ - n \Delta z},
\label{phi2D}
\end{equation}
\begin{equation}
\Delta z = z - \eta_0,\;
\eta(x,t) = \eta_0(t) + \sum_{n=1}^{N} K_n(t) x^{2n}/(2n)!,
\label{eta2D}
\end{equation}
where $c_{nx}=\cos nx$ and $N$ is the truncation number.
It defines the order of approximation of conditions (\ref{continuity},\ref{bernouilli}).

{\it 2D solution.}
The expansion (\ref{phi2D}) satisfies $\Delta\varphi=0.$
The expressions (\ref{phi2D},\ref{eta2D}) introduce geometrical $\eta_0, K_1,..,K_N$
and velocity $\varphi_1,..,\varphi_N$ unknowns
for the ordinary differential equation system
in our method of asymptotic collocations (MAC).
We say asymptotic collocations
because of
the close connection to the method of ordinary collocations
in which boundary conditions are approximated
in a set of discrete points $\{x_i\}.$
The equations of the MAC
appear asymptotically
when all points $x_i$ tend to point B, or S, or J\cite{INA,INA2}.
In 2D,
these equations,
for $N\leq 6,$
were first derived
and integrated in\cite{INA2}
in case of bubbles.

The Fig. \ref{fig-ripples} presents the $N=5$ solution.
Here, for the first time,
we use high-order MAC
to study jets.
We carefully describe acceleration of the jet
and for the RM case
obtain very accurate values
for its terminal (asymptotic) velocity,
$w_J(t=\infty, N=5)=-1.923$
$(w_J(t)\equiv\dot\eta_0\equiv\dot\eta (x_J,t)),$
standard initial conditions.
The accuracy
of this value
is
$\varepsilon(1)=10^{-1.0},$
$\varepsilon(2)=10^{-2.3},$
$\varepsilon(3)=10^{-3.16}$
and $\varepsilon(4)=10^{-3.43},$
where $\varepsilon (N)=
|w_J(\infty, N+1)-w_J(\infty, N)|/|w_J(\infty, N)|.$
Since the error is very small,
the MAC may be used to control the accuracy
of other methods.

{\it 3D solution.}
For the B4 lattice we have
$$
\varphi^{(B4)}(x,y,z,t) = \sum \sum\varphi_{nm}(t)
c_{nx} c_{my} e^{(4)}_{nm},
$$
$$
\eta(x,y,t) = \sum \sum K_{nm}(t) x^{2n} y^{2m}/(2n)!(2m)!,
$$
where
$0 \leq n+m \leq N,$
$e^{(4)}_{nm} = \exp ( - q^{(4)}_{nm} \Delta z),$
$K_{00}(t)=\eta_0(t)$
and
$q^{(4)}_{nm}=\sqrt{n^2+m^2}.$
The unknowns in the system are $\eta_0, K_{nm}, \varphi_{nm}.$
For $N=1,$ the indices $(n,m)$ are $10$ and $01.$
For $N=2,$ they are $10, 01, 20, 11$ and $02.$
Points B and J
(but not S)
are symmetric,
$\varphi_{nm}=\varphi_{mn}$ and $K_{nm}=K_{mn}.$
At the lowest order $N=1$
(Layzer approximation\cite{INA,INA2,Layzer,Shv,Mik,Ab,Zhang})
the unknowns,
at points B and J,
are $\eta_0,K,w,$
where $K=K_{10}=K_{01}$ is the curvature
and $w=-\varphi_{10}=-\varphi_{01}$ is the velocity of a bubble or a jet.
The system $N=1$ for the B4 lattice,
valid in points B and J,
has been derived
and
solved
for the RM case in \cite{INA2}.
The 2D $N=1$ system
was considered in \cite{INA,INA2,Layzer,Shv,Mik,Ab,Zhang}
and 3D $N=1$ systems for B6 and B4 cases were examined
in \cite{INA,INA2,Ab}.

This is the first time we use MAC
to describe the dynamics
of all three kinds of points (B, S and J) at order $N=2.$
It appears much more complicated
than $N=1$ (compare also with 2D, $N\leq 5).$
The terminal velocities of RM jet and saddle are
$w_J(\infty,N=1)=-\sqrt{2},$
$w_J(\infty,N=2)=-1.698$
and
$w_S(\infty,N=1)=-0.572,$
$w_S(\infty,N=2)=-0.512.$
This gives an accurate asymptotic value
of the geometrical ratio (\ref{gamma}),
$\Gamma\to w_S(\infty)/w_J(\infty)$
as $t\to\infty.$
Fig. \ref{fig-bubbles} gives an example of second order solution.

For the B6 and B3 lattices,
we have in first order
$$
\varphi^{(B6,3)} = 
[\varphi^{+0}(t) c^+ +
 \varphi^{-0}(t) c^- +
 \varphi^{+-}(t) c] e^{-\Delta z},
$$
where $c^{\pm}=\cos \vec k^{\pm}\vec r,$ $ c=\cos \vec k^0\vec r,$
$\vec k^{\pm} = \{1/2, \pm\sqrt{3}/2, 0\},$
$\vec k^0=\{1,0,0\}$
and $\varphi^{+0}=\varphi^{-0}=\varphi^{+-}$
for B and J points.


The first order dynamical systems for points B and J
for all three lattices B6, 4 and 3
are the same\cite{INA2}:
$
\dot\eta_0=w,
$
$
\dot w = - [ w^2 + 4 g(t) K ]/2 ( 1 + 2K ),
$
$
\dot K = - ( 1 + 4K ) w/2,
$
where
$
\dot f\equiv df/dt.
$
In this system,
we emphasize that
$g(t)$
is an arbitrary function.
Eliminating $t$
between first and last equations,
we obtain
$dK/d\eta_0 = - 1/2 - 2K.$
The solution is
\begin{equation}
K(\eta_0)=-1/4 + \exp ( - 2 \eta_0 )/4.
\label{order1_sol}
\end{equation}
The 2D analog is
$dK/d\eta_0 = - 1 - 3K$
and
$K=-1/3 + \exp ( - 3 \eta_0 )/3.$
The solution (\ref{order1_sol}) has linear asymptotes for $|\eta_0|\ll 1$
and tends to $-1/4$
for $|\eta_0|\gg 1.$
In the linear stage we have
$\eta \approx \eta_0(c_{1x}+c_{1y})/2$ (B4),
$\eta \approx \eta_0(c^+ +c^- +c)/3$ (B6,3)
and $K \approx - \eta_0/2.$

In RM case, the differential system has integrals \cite{INA2}
\begin{equation}
\sqrt{\frac{1+2K}{1+4K}}-
\frac{1}{\sqrt{2}}
\ln\frac{\sqrt{2+4K}+\sqrt{1+4K}}{(\sqrt{2}+1)\exp(-\sqrt{2})}=w_0 t,
\label{integral1}
\end{equation}
\begin{equation}
\frac{w_0}{w}-1+
\frac{1}{2\sqrt{2}}
\ln\left(
\frac{\sqrt{2}+1}{\sqrt{2}-1}
\frac{\sqrt{2}w_0-w}{\sqrt{2}w_0+w}
\right)=w_0 t.
\label{integral2}
\end{equation}
Substituting solution $K(\eta_0)$ (\ref{order1_sol}) into (\ref{integral1})
we obtain
\begin{equation}
\sqrt{\frac{1+e^{2\eta_0}}{2}}-1-
\frac{1}{\sqrt{2}}\ln\frac{\sqrt{1+e^{-2\eta_0}}+e^{-\eta_0}}{\sqrt{2}+1}=
w_0 t.
\label{firstorder}
\end{equation}
In (\ref{integral1}-\ref{firstorder}) $w_0$ is the initial velocity
of B or J points.
We will assume
that this initial velocity
for the tip of bubble B is equal to 1
for all lattices.
Then initial velocities of points J are
$-1/2,-1$ and $-2$ for lattices B6,4 and 3
respectively.
Similarly to (\ref{integral1}-\ref{firstorder}),
the 2D solution has been obtained previously\cite{INA,INA2,Mik,Zhang}.
The integrals (\ref{integral1}-\ref{firstorder}) give
$\eta_0(t),K(t),w(t)$
in analytic form for all times
from initial to asymptotic state.
They are valid in the RM case
for B and J points
of B6,4 and 3 lattices.

It is very suprising
that the relation $K(\eta_0)$ (\ref{order1_sol})
between the main geometrical characteristics
$\eta_0$ and $K$
for a bubble
penetrating into the dense fluid
does not depend upon $g(t)$
for $N=1.$
This means that the relation is only weakly dependent on $g(t)$
in the general case with arbitrary $N.$

The system for $N=2$ is rather long
and can not be given here.
However,
for saddles and $N=1,$
we have:
\begin{equation}
\dot\eta_0=\alpha-\beta,
\label{2Seta4}
\end{equation}
\begin{equation}
\dot K=-(1+3K)\alpha+K\beta,
\label{2SK4}
\end{equation}
\begin{equation}
\dot Q=-Q\alpha+(1+3Q)\beta,
\label{2SQ4}
\end{equation}
\begin{equation}
-(1+K)\dot\alpha+K\dot\beta=\alpha^2+g(t) K,
\label{2SbernouilliK4}
\end{equation}
\begin{equation}
-Q\dot\alpha+(1+Q)\dot\beta=\beta^2+g(t) Q
\label{2SbernouilliQ4}
\end{equation}
in the B4 case with
$\varphi=(-\alpha c_{1x}+\beta c_{1y})e^{-\Delta z}.$
For B6,3 we have
\begin{equation}
\dot\eta_0=2\alpha-\gamma,
\label{2Seta63}
\end{equation}
\begin{equation}
2\dot K=-(1+6K)\alpha+2(1+3K)\gamma,
\label{2SK63}
\end{equation}
\begin{equation}
2\dot Q=-(3+10Q)\alpha+2Q\gamma,
\label{2SQ63}
\end{equation}
\begin{equation}
-2(1+4K)\dot\alpha+4(1+K)\dot\gamma=(\alpha-2\gamma)^2+4g(t) K,
\label{2SbernouilliK63}
\end{equation}
\begin{equation}
-2(3+4Q)\dot\alpha+4Q\dot\gamma=9\alpha^2+4g(t) Q,
\label{2SbernouilliQ63}
\end{equation}
where $\alpha$ and $\gamma$ are the amplitudes of potential
$\varphi=[-\alpha(c^++c^-)+\gamma c]e^{-\Delta z}$
written with respect to the point S.
In the standard case,
initial conditions are
$\alpha(0)=\gamma(0)=-1/3$ (B6),
$\alpha(0)=\beta(0)=-1/2$ (B4),
$\alpha(0)=\gamma(0)=2/3$ (B3)
and
$\eta_0(0)=K(0)=Q(0)=0$ (B6,4,3).
From (\ref{2Seta4}) and (\ref{2Seta63}),
initial velocities of the saddles are
$-1/3$ (B6), $0$ (B4) and $2/3$ (B3) -
to be compared with initial velocities of jets.
Although,
the systems are the same
for B6 and B3,
the initial conditions differ.
From systems (\ref{2Seta4}-\ref{2SbernouilliQ4}) or (\ref{2Seta63}-\ref{2SbernouilliQ63})
we found numerically the trajectories of saddles $\eta_0(t)$
and the evolution of the curvatures $K(t), Q(t).$
The terminal velocities of saddles in RM case are
$w_S(\infty,N=1)=-0.748$ (B6),
$w_S(\infty,1)=-0.572$ (B4)
and
$w_S(\infty,1)=-0.196$ (B3).

In Fig. \ref{fig-ratio} we present the time variation of the geometrical ratio $\Gamma$.
Initial values of the ratio are
$\Gamma_{B6}(t=0)=8/9,$
$\Gamma_{B4}(0)=1/2$
and
$\Gamma_{B3}(0)=1/9.$
Simulations $(\mu=0.1)$ fit rather well with the theory $(\mu\to 0)$
although $\mu$ differs.
We observe
that the agreement between theory and simulation
is better
for the RMI than for the RTI.
The increase of $N$ significantly improves the accuracy
(for the B4 system, the $N=2$ curve and the one coming from
simulation are very close for the RMI). 
The function $g(t)$ influences therefore the evolution of this ratio.
The morphology of the ejecta mainly depends on the type of lattice.
The shortest skirts are obtained for the B3 lattice (lower set of curves)
which produces powerful and fast jets (Fig. 3).
Previously, it has been shown \cite{INA2}
that for random (turbulent) cases,
the patterns are similar to the B6 and B4 lattices
(B3 lattices occur only for special conditions).
Moreover, B3 type structures may appear after reshock
and the corresponding rephasing because B6 bubbles transform into B3 jets.

In summary, we have considered the effects of lattice and
time-dependant acceleration
on the evolution of the interface.
We found that the shape of imprinting bubbles is very weakly dependent
on these factors,
but the position of the bubbles depends on the history of the acceleration.
At the same time, the shape and dynamics of the ejecta appear very sensitive
to both factors.

The authors would like to thank
J.F. Haas
for many interesting discussions and Douglas Wilson for careful reading.
A.M.O. and N.A.I. are grateful for the support
from
RBRF (00-01-00250, 99-02-16666)
and
from Landau Institute - CNRS jumelage (N.A.I.).

      \begin{figure}[htbp]
      \caption{Effect of lattice symmetry and acceleration history on
      the evolution of the shape of $\eta.$ The upper, middle and
      lower sets of curves correspond respectively to the B6, B4 and B3
      lattices. The grey and black  curves correspond respectively to
      numerical simulations - see Fig. 3 - and to the analytical
      approach with $N=1$. The dashed curves have been obtained from
      theory with $N=2$. The curves labelled by 1 (resp. 2) describe
      the RTI (resp. RMI).}
      \label{fig-ratio}
      \end{figure}

\end{document}